\documentclass[aps,twocolumn,groupedaddress,amssymb]{revtex4-1}

\usepackage{latexsym}
\usepackage{amsbsy}
\usepackage{amsmath}
\usepackage{graphicx}
\usepackage{xcolor}
\usepackage{bm}

\usepackage[colorlinks=true]{hyperref}
\hypersetup{
    bookmarks=true,
    pdftoolbar=true,
    pdfmenubar=true,
    pdfstartview={FitH},
    pdftitle={Altermagnetic Superconducting Diode Effect in Mn3Pt/Nb Heterostructures},
    pdfauthor={Sachin et al.},
    colorlinks=true,
    linkcolor=blue,
    citecolor=blue,
    urlcolor=blue
}

\newcommand{\figref}[1]{Fig.~\ref{#1}}

\begin{document}

\title{Altermagnetic Superconducting Diode Effect in Mn$_{3}$Pt/Nb Heterostructures}

\author{Saurav Sachin\textsuperscript{1}}
\author{Mathias S. Scheurer\textsuperscript{2}}
\altaffiliation{Contact author: mathias.scheurer@itp3.uni-stuttgart.de} 
\author{Constantin Schrade\textsuperscript{3}}
\altaffiliation{Contact author: cschrade@lsu.edu} 
\author{Sujit Manna\textsuperscript{1}}
\altaffiliation{Contact author: smanna@physics.iitd.ac.in} 

\affiliation{\textsuperscript{1}Department of Physics, Indian Institute of Technology Delhi, Hauz Khas, New Delhi 110016, India}
\affiliation{\textsuperscript{2}Institute for Theoretical Physics III, University of Stuttgart, 70550 Stuttgart, Germany}
\affiliation{\textsuperscript{3}Hearne Institute of Theoretical Physics, Department of Physics and Astronomy, Louisiana State University, Baton Rouge, Louisiana 70803, USA}

\begin{abstract}
Compensated magnetic orders that can split the spin-degeneracy of electronic bands have become a very active field of research. As opposed to spin-orbit coupling, the splitting resulting from these “altermagnets” is not a small relativistic correction and, in contrast to ferromagnets, not accompanied by a net magnetization and large stray fields. In particular, the theoretical analysis of the interplay of altermagnetism and superconductivity has taken center stage, while experimental investigations of their coexistence remain in their infancy. We here study heterostructures consisting of Nb thins films interfaced with the $T_1$ and $T_2$ phases of Mn$_3$Pt. These non-collinear magnetic states can be thought of as descendants from the same altermagnetic order in the absence of spin-orbit coupling. We demonstrate the non-trivial impact on the superconducting state of Nb, which exhibits a zero-field superconducting diode effect, despite the compensated ($T_2$) and nearly-compensated ($T_1$) magnetic order; the diode efficiencies can reach large values (up to 50$\%$). The diode effect is found to be highly sensitive to the form of the magnetic order, illustrating its potential as a symmetry probe. The complex magnetic field and temperature dependence hint at a rich interplay of multiple contributing mechanisms. Our results define a new materials paradigm for dissipationless spintronics and magnetization-free diode functionality, while motivating further exploration of non-collinear altermagnetic superconductors.
\end{abstract}

\maketitle

\section{Introduction}
Although initially motivated by its relevance to spintronics, altermagnetism has become a very active interdisciplinary field \cite{PhysRevX.12.040501,Fukaya_2025,Review3,magnetism5030017,PhysRevX.12.040002,RafaelReview}. The defining features of altermagnets are (i) the presence of symmetries that guarantee a vanishing net magnetization (like in antiferromagnets) while (ii) inducing a finite spin-splitting of the electronic bands already in the limit of zero spin-orbit coupling (like ferromagnets). Initially, altermagnetism was defined to refer exclusively to collinear magnetic orders \cite{PhysRevX.12.040501}. However, the defining properties mentioned above can also be applied to non-collinear magnetic textures, as recognized in multiple recent works
\cite{PhysRevMaterials.5.014409,PhysRevB.101.220403,NonColl1,NonColl2,NonColl3,PhysRevB.110.214428,SpinSplittingExp,PhysRevB.109.024404,banerjee2024altermagnetic}.

Most recently, there has been growing interest in the interplay of pairing and altermagnetism \cite{Fukaya_2025}. 
A particularly promising direction is the proposal\cite{banerjee2024altermagnetic} that altermagnets can stabilize a superconducting diode effect (SDE) without net magnetization or external magnetic fields.
The SDE, which refers to an asymmetry of the critical current in opposite directions, is of interest as it could provide a route to dissipation-less rectification enabled by broken inversion and time-reversal symmetries \cite{DiodeReview1,DiodeReview2}.
An altermagnetic implementation of the SDE is particularly appealing as it is not associated with detrimental stray fields.
This property differentiates the SDE in altermagnets from previous implementations that usually require an external magnetic field or an effective
ferromagnetic order with a finite magnetic moment\cite{Lin2022,FerromagneticProximity,le2024superconducting,wu2022fieldfree}. 
While the SDE in altermagnets has been explored theoretically by several groups\cite{b7rh-v7nq,2025PhRvL.135b6001C}, its experimental realization is still an important open problem.  

In general, experiments on altermagnetic superconductivity are at an early but rapidly developing stage\cite{Fukaya_2025}. 
For example, there are altermagnetic candidate materials that have been demonstrated to host superconductivity\cite{PhysRevLett.125.147001,FeSeAM,PhysRevX.12.031042}. Moreover, in a planar junction of a superconducting In electrode and the altermagnetic candidate MnTe, signs of Andreev reflection have been reported. 
In addition, signatures of triplet pairing have been observed\cite{Parkin1,Parkin2} in Josephson junctions involving superconducting Nb leads coupled to Mn$_3$Ge, which is a candidate material for noncollinear altermagnetism\cite{NonColl3}.

\begin{figure*}
    \centering
    \includegraphics[width=1\linewidth]{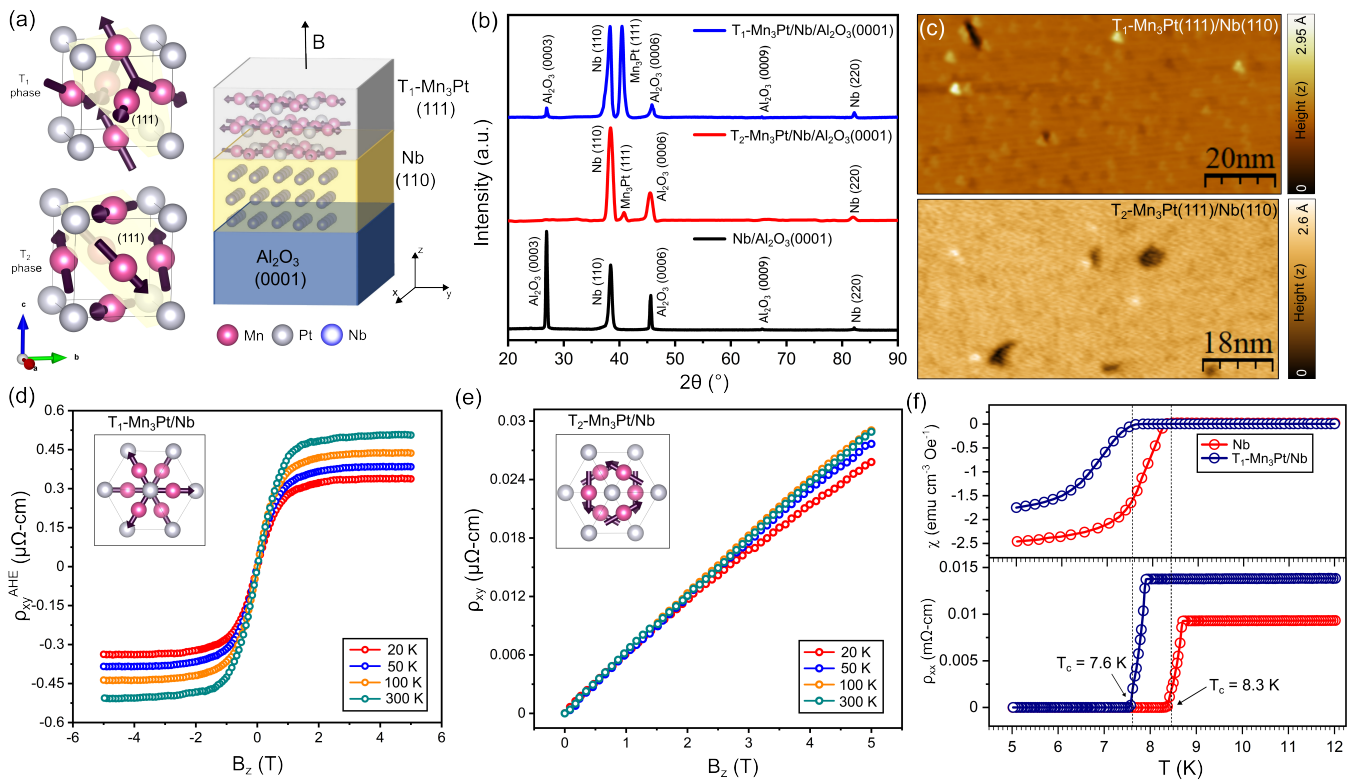}
    \caption{\textbf{Noncollinear magnetism in Mn$_3$Pt and superconducting Nb films interface}: (a) Schematic representation of the spin configurations in cubic Mn$_3$Pt: $T_1$ phase exhibits an all-in/all-out noncollinear spin arrangement on the Mn sublattice, while the $T_2$ phase shows a head-to-tail configuration. Illustration of the Mn$_3$Pt(111)/Nb(110) heterostructure epitaxially grown on a Al$_2$O$_3$(0001), with field applied perpendicular to the film. (b) X-ray diffraction patterns of a 25 nm Nb film and $T_1$ ($T_2$)-Mn$_3$Pt (15 nm)/Nb (25 nm) bilayer. (c) Constant-current STM topography (100 nm $\times$ 50 nm) obtained on $T_1$-Mn$_3$Pt(111)/Nb(110)/Al$_2$O$_3$(0001) and $T_2$-Mn$_3$Pt(111)/Nb(110)/Al$_2$O$_3$(0001) (bottom panel, 90 nm $\times$ 40 nm)  films. The measurements were performed under tunneling conditions of I$_t$ = 0.2 nA, U$_{\text{bias}}$ = 0.5 V and I$_t$ = 1 nA, U$_{\text{bias}}$ = 1 V respectively. (d,e) Magnetic field dependence of Hall resistivity ($\rho_{xy}^{\text{AHE}}$) for (d) $T_1$-Mn$_3$Pt/Nb and (e) $T_2$-Mn$_3$Pt/Nb bilayers. We observe nonvanishing Berry curvature-induced large anomalous Hall response in $T_1$-Mn$_3$Pt and negligible Hall signal in $T_2$-Mn$_3$Pt. This represents strong phase dependence of anomalous Hall resistivity in Mn$_3$Pt. (f) Magnetic susceptibility ($\chi$) versus temperature near the superconducting transition temperature ($T_c$), alongside the temperature-dependent longitudinal resistivity ($\rho_{xx}$) of both Nb film and $T_1$-Mn$_3$Pt/Nb heterostructure. No significant change in $T_c$ is observed between the $T_1$ and $T_2$ phases.}
    \label{fig:1}
\end{figure*}

Motivated by this progress in the field, we here provide the first experimental evidence of a SDE at zero external field in a heterostructure of superconducting Nb proximity-coupled to the noncollinear altermagnet Mn$_3$Pt. Mn$_3$Pt is known to host several energetically close magnetic states, including the $T_1$ and $T_2$ \cite{kren1968magnetic, rimmler2023atomic, PhysRevB.92.144426,rimmler2025non} phases relevant to us here. As can be seen in the schematic illustration in the left two panels of \figref{fig:1}(a), these two non-collinear orders have their spins (mainly) aligned in the kagome $(111)$ plane with a relative angle of $120^\circ$ between neighboring spins. Their near degeneracy immediately follows from the fact that they are related by a global spin rotation and, hence, are equivalent in the absence of spin-orbit coupling. The (111) surface of Mn$_3$Pt in our heterostructure still retains a point group $C_{3v}$, generated by three-fold rotation $C_{3z}$ along the $(111)$ direction and mirror operations $\sigma_v$. Without spin-orbit coupling, the combination of three-fold rotation in spin and real space, $C^s_{3z}$, is still a symmetry, leading to the vanishing of the in-plane magnetization; furthermore, $\sigma_v$ combined with a two-fold spin rotation along the direction of the spins of the sites on the mirror plane prohibits any out-of-plane magnetization. However, the heterostructure has no symmetries left which would protect the spin degeneracy of the bands, see also \cite{PartnerTheoryPaper,PhysRevResearch.2.033112,PhysRevB.99.165141,PhysRevLett.112.017205}. As such, the magnetic order at the interface layer can be thought of as a non-collinear altermagnet \cite{NonColl2,NonColl3}. Due to the broken inversion symmetry, the spin splitting is not entirely even in momentum, which---depending on conventions---can be thought of as an admixed ``antialtermagnetic'' component \cite{RafaelReview}. For simplicity, we here follow \cite{NonColl2} and refer to the magnetic order simply as an altermagnet (see our companion theory paper \cite{PartnerTheoryPaper} for more details).

Turning on spin-orbit coupling in this parent altermagnetic phase makes the $T_1$ and $T_2$ orders distinct. While both preserve $C^s_{3z}$, only $T_2$ keeps the spinful reflection symmetry $\sigma_v^s$ such that it remains symmetry-compensated even with spin-orbit coupling. In contrast, the $T_1$ state breaks this symmetry; as a result, it is chiral and expected to develop a small but finite canting \cite{fruchart1977magnetic} and out-of-plane magnetization---in agreement with \cite{PhysRevB.92.144426} where it was found to be only of order of $1 \%$.

Importantly, the $T_2$ state realizes one of the very few combinations of point groups and magnetic order parameters \cite{banerjee2024altermagnetic} with fully compensated magnetization where a SDE follows by symmetry. Meanwhile, the $T_1$ state can be thought of as a related state, emerging from the same parent altermagnetic order, with a small spin-orbit-coupling-induced canting; by symmetry, it is also expected to induce a SDE. In agreement with these expectations, we observe a SDE for both $T_1$ and $T_2$ orders which clearly demonstrates the impact of magnetism on the proximitized superconductor (Nb). Along the direction we probe within our transport study, the efficiency is significantly larger for the $T_1$ state and can be controlled to a large degree with an applied magnetic field.

\section{Methods}

Nb and Mn$_3$Pt thin films were grown in an ultrahigh vacuum (UHV) magnetron sputtering system with a base pressure of $<$ 2 $\times$ 10$^{-8}$ mbar \cite{Sinha2025_2}. Polished sapphire Al$_2$O$_3$ (0001) substrates were prepared by standard exsitu chemical cleaning and subsequent thermal annealing up to 600\textdegree C for repeated cycles in UHV until monoatomic steps form in the surface. Nb films were grown using Nb source (99.99\% purity) at 650\textdegree C in an argon environment with a working pressure close to 1.5 $\times$ 10$^{-2}$ mbar. Different compositions of thin films of Mn$_{3+x}$Pt$_{1-x}$ \cite{sinha2025} were prepared by co-deposition of high-purity Mn and Pt sources on epitaxial Nb films. The structural characterization of the films was performed using X-ray diffraction (XRD), using a PANalytical X’Pert system with a Cu-K$_{\alpha}$ radiation source ($\lambda$ = 1.5418 Å). X-ray reflectivity (XRR) measurements were used to quantify the film thickness and growth rate. The surface morphology was characterized using atomic force microscopy (AFM) (Oxford Instruments Asylum Research, MFP-3D system). Scanning tunneling microscopy (STM) measurements were performed at room temperature using a customized STM operated by commercial controller with active vibration cancellation setup \cite{sinha2025}. STM tips were prepared from polycrystalline PtIr by chemical etching followed by in-situ high-temperature annealing. Before STM measurements, the samples were also annealed in vacuum at 300\textdegree C to remove surface adsorbates and contaminants acquired during transport from the growth chamber to the STM chamber. Electrical contact between the sample and sample holder was established using high conducting glue. All imaging were performed in constant current mode. Standard photolithography and Ar ion milling were used to pattern the films into Hall bar geometry with width $\sim$ 10 $\mu$m and length $\sim$ 200 $\mu$m, followed by gold (Au) contact deposition. Electrical transport measurements down to 2K were performed using physical property measurement system (PPMS). Four-probe technique was employed to measure current-voltage (I-V) characteristics. The measurements were taken at various fixed temperatures with stability within $\pm$2 mK. Magnetic measurements down to 2K were performed using Quantum Design magnetic property measurement system (MPMS3) respectively. The magnetic field was applied perpendicular to the sample plane (out-of-plane) during magneto-transport measurements.

\section{Results}

\begin{figure*}
    \centering
    \includegraphics[width=1\linewidth]{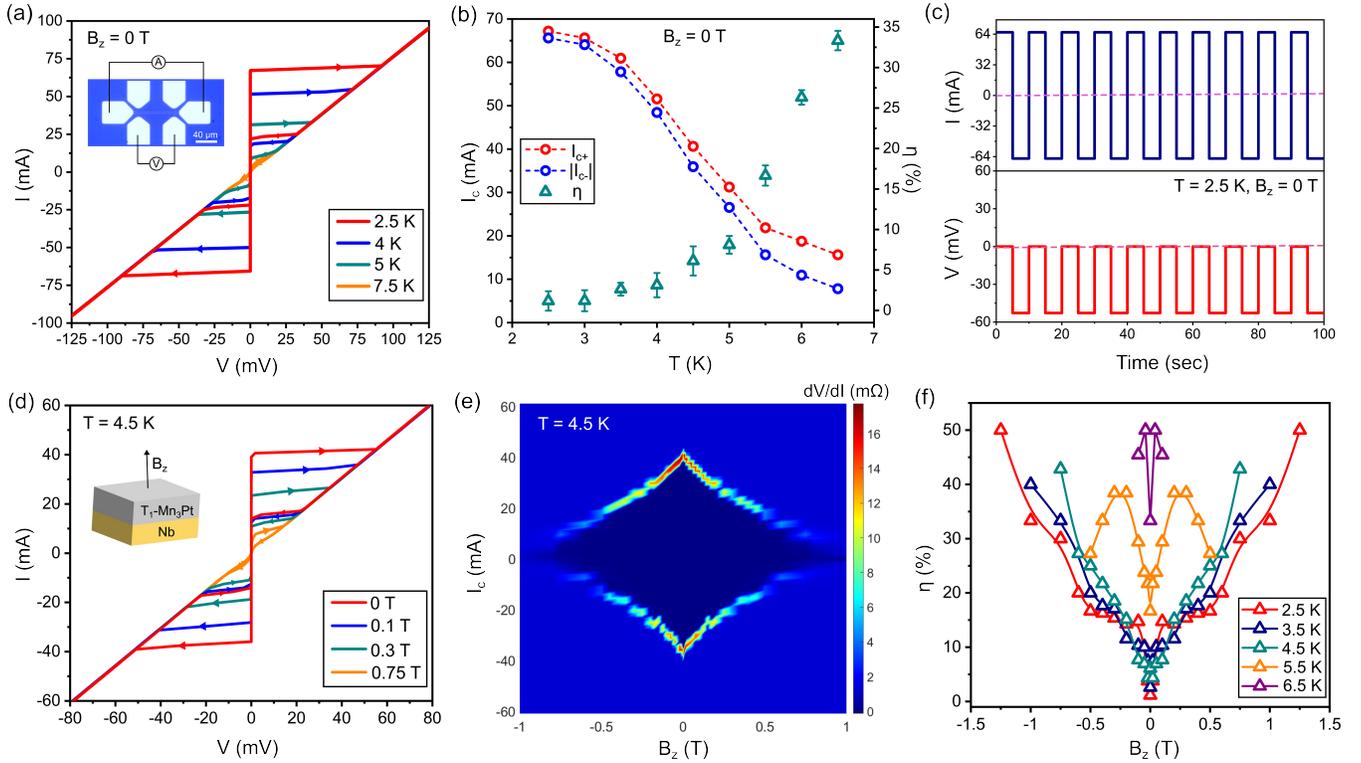}
    \caption{\textbf{Giant SDE in $T_1$-Mn$_3$Pt/Nb}: (a) Current-voltage (I-V) characteristics of $T_1$-Mn$_3$Pt/Nb measured in zero magnetic field over a temperature range of 2.5 - 7.5 K. Optical microscope image of Hall bar strip (inset) of Mn$_3$Pt/Nb heterostructure, in which scale bar denotes 40 $\mu$m. (b) Temperature dependence of absolute value of nonreciprocal critical current and corresponding diode efficiency at zero field. We observe pronounced hysteresis in I-V curves, while the SDE peaks slightly below $T_c$ and vanishes at superconducting transition temperature. Diode efficiency reaches its maximum around 33\% at 6.5 K in absence of any external field. (c) Demonstration of zero-field supercurrent rectification at 2.5 K via periodic switching between superconducting and normal conducting states by changing polarity of applied currents between $I_c^{+}$ and $\left| I_c^{-} \right|$. (d) I-V characteristics of $T_1$-Mn$_3$Pt measured under different out-of-plane magnetic fields ($B_z$). Note that SDE does not depend on the polarity of magnetic field. (e) Contour map showing evolution of differential resistance versus critical current and magnetic field at 4.5 K. The plot highlights oscillatory modulation of critical supercurrents with magnetic field, indicating of interference effects arising from superconducting order parameter. (f) Evolution of diode efficiency with magnetic field ($B_z$) at different temperatures. The data reveals that the SDE is further amplified by the application of external magnetic fields, with the enhancement being strongly temperature dependent.}
    \label{fig:2}
\end{figure*}

The schematic illustration of growth of $T_1$-Mn$_3$Pt with (111) orientation on Nb(110), deposited on a sapphire (Al$_2$O$_3$) substrate, is depicted in the right panel of \figref{fig:1}(a). Figure \ref{fig:1}(b) shows the representative x-ray diffraction (XRD) patterns of the Nb thin film, with and without $T_1$ and $T_2$-Mn$_3$Pt/Nb bilayer. The Nb layer exhibits a body-centered cubic (BCC) structure with calculated lattice parameter of a = 3.31 Å, consistent with bulk Nb. The observed diffraction peaks confirm the crystallinity and phase purity of the films and peak positions are consistent with previous reports \cite{du2016nbinterface, sinha2025}. The Mn$_3$Pt (111) peak position shifts to slightly lower diffraction angle for Mn-rich films ($T_1$-Mn$_3$Pt) due to epitaxial strain. Our structural and magnetic characterizations, including surface morphology, film quality, epitaxial growth relation, longitudinal resistivity and magnetization measurements, are presented in the supplementary information \cite{SuppRef1}. Further, we characterized the surface topography of Nb and Mn$_3$Pt/Nb films using constant-current scanning tunneling microscopy (STM). The STM differential topograph of Nb(110) films grown on Al$_2$O$_3$(0001) is presented in supplementary information (Fig.~S2), which reveals atomically flat terraces with minimal roughness. STM topography of Mn$_3$Pt(111)/Nb(110) (shown in \figref{fig:1}(c)) shows the presence of atomically sharp, extended terraces with a monolayer-high underline step (black depressions) confirming high-quality epitaxial growth of Mn$_3$Pt(111) with a clean surface and well-defined interface to the Nb film. STM further ensures similar mode of growth for both T$_1$ and T$_2$ phases of Mn$_3$Pt.

\begin{figure*}
    \centering
    \includegraphics[width=1\linewidth]{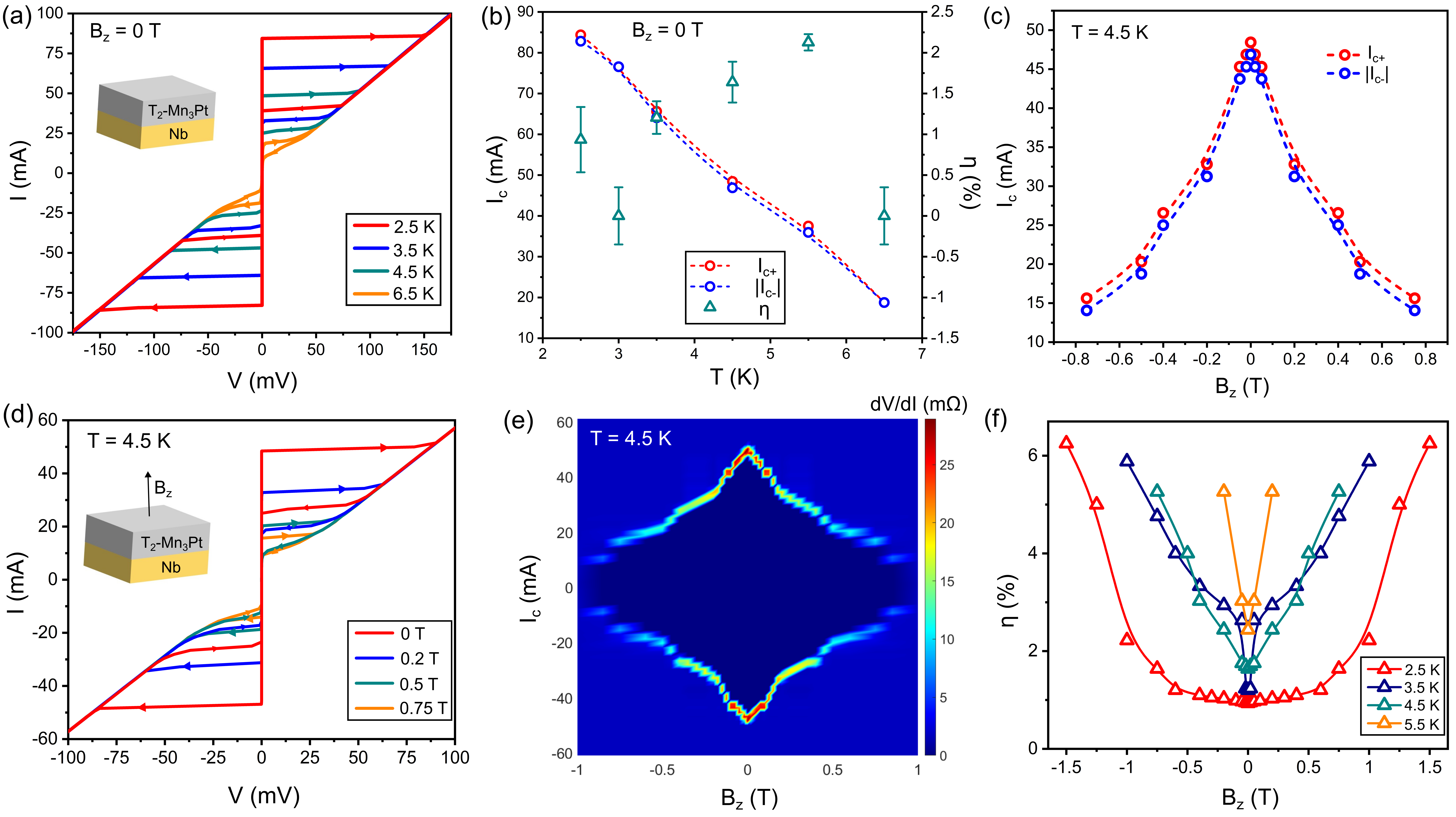}
    \caption{\textbf{Weak SDE in $T_2$-Mn$_3$Pt/Nb}: (a) Zero-field I-V characteristics of $T_2$-Mn$_3$Pt/Nb measured across temperature range from 2.5 K to 6.5 K showing a reduced rectification with negligible asymmetric critical currents. (b) Temperature dependence of positive and negative critical currents along with corresponding zero-field diode efficiency, plotted as a function of temperature. (c) Variation of critical currents with application of magnetic field at 4.5 K.(d) Magnetic field dependence of I-V characteristics for $T_2$-Mn$_3$Pt/Nb at 4.5 K. (e) Symmetric dV/dI mapping as a function of current bias and magnetic field at 4.5 K. (f) Diode efficiency at different temperatures as a function of magnetic field, which is non-zero as expected by symmetry for the $T_2$ state but much weaker than in the $T_1$-Mn$_3$Pt/Nb counterpart.}
    \label{fig:3}
\end{figure*}

In our recent works \cite{sinha2025, Sinha2025_2}, we demonstrated that the two magnetic phases, the $T_1$ and $T_2$ state defined above, can be stabilized by varying the compositions $x$ in Mn$_{3+x}$Pt$_{1-x}$ via co-sputtering. Higher Mn concentration favors growth of the chiral $T_1$ phase. To verify the presence of the two different magnetic phases in our heterostructures involving Nb, we focus on temperatures above the superconducting transition temperature $T_c$. The transverse Hall resistivity as a function of magnetic field measured on $T_1$- and $T_2$-Mn$_3$Pt/Nb is shown in \figref{fig:1}(d,e). While sweeping the magnetic field, the contribution of the anomalous Hall effect (AHE) to the transverse Hall resistivity can be easily distinguished between both magnetic phases. $T_1$-Mn$_3$Pt ($x = 0.28$) exhibits a pronounced Hall response arising from its chiral nature, which allows for a non-vanishing Berry curvature \cite{PhysRevLett.112.017205}. In contrast, in the $T_2$ phase, the presence of $\sigma_v^s$ prohibits a net Berry curvature contribution; this is in line with our measurements with much smaller $\rho_{xy}$ in $T_2$-Mn$_3$Pt ($x = 0.09$).

Building on this understanding, we will now focus on the SDE in the $T_1$ and $T_2$ phases of Mn$_3$Pt/Nb, arising from the interplay between non-trivial magnetic textures and superconductivity.
To characterize the superconducting properties, we performed resistivity measurements as a function of temperature. We estimated the superconducting critical temperature ($T_c$) of the Nb film to be 8.3 K as shown in \figref{fig:1}(f), which is close to the previously reported values \cite{tanatar2022nbdisorder,patel2019nbti}. Upon interfacing with Mn$_3$Pt, $T_c$ of the Nb layer is noticeably suppressed to 7.6~K, characterized by magnetic susceptibility and electrical resistivity measurements. This suppression provides our first indication of a non-trivial effect of the magnetic order on the superconducting state. Indeed, due to the broken time-reversal symmetry, altermagnetism is generally expected to suppress the critical temperature of pairing \cite{PhysRevB.111.L100502,PhysRevB.108.054510,2025arXiv251019943S}. 

To probe the interplay of magnetism and superconductivity more directly, we systematically performed current-voltage measurements as a function of temperature in the absence of a magnetic field and investigated the SDE.
We measured I–V characteristics in a two-step sweep process: positive sweep, where the DC current was ramped from zero to a positive maximum and then back to zero ($0 \rightarrow p_{\text{max}} \rightarrow 0$), followed by negative sweep from zero to negative maximum and subsequently return to zero ($0 \rightarrow n_{\text{max}} \rightarrow 0$). The critical current $I_c$ is defined as the current at which the junction switches to the normal state, while the retrapping current $I_r$ corresponds to the current for returning back to the superconducting state (see Fig.~S5 for more details). 

We start our discussion with $T_1$-Mn$_3$Pt/Nb, with results presented in \figref{fig:2}. The I–V characteristics, shown in \figref{fig:2}(a), reveal a pronounced superconducting state from 2.5 K to 6.5 K, characterized by sharp voltage jumps at critical currents. The interface shows a large hysteresis behavior at lower temperatures, which indicates an underdamped regime, where capacitive effects dominate over dissipation in the phase dynamics.
As the temperature increases, the magnitude of the critical current gradually decreases, approaching zero near the superconducting transition temperature. This can also be more directly seen in \figref{fig:2}(b) where we plot the critical current along the positive ($I_{c+}$) and negative ($I_{c-}$) directions. While both approach zero, as $T$ approaches $T_c$ from below, they significantly differ in magnitude, revealing the presence of a SDE. As the latter requires broken time-reversal symmetry, this provides direct evidence of the impact of the magnetic order of Mn$_3$Pt on the superconducting state of Nb. As already pointed out above (see also \cite{PartnerTheoryPaper}), the symmetry reduction associated with the $T_1$ phase is sufficient to stabilize the SDE, without any external magnetic field.

To quantify the SDE, we estimate the degree of nonreciprocity via the efficiency, $\eta = (I_{c+} - \left| I_{c-} \right|)/(I_{c+} + \left| I_{c-} \right|)$. This is also plotted in \figref{fig:2}(b) revealing that $\eta$ rises sharply above 4 K. It reaches a maximum near $T_c$, suggesting an enhanced diode efficiency close to the superconducting transition and reaching values on the order of $50\%$. 
The distinct values of $I_{c+}$ and $\left| I_{c-} \right|$ imply that when we apply a current ranging between these values, the system behaves like a superconducting state for current flow in one direction and like a normal resistive state for the opposite direction. This directional dependence manifests as a superconducting rectification effect (see \figref{fig:2}(c) for rectification at 2.5~K) that occurs for current magnitudes smaller than $I_{c+}$, but larger than $\left| I_{c-} \right|$.

Having established a zero-field SDE, we further measured I-V curves of upward and downward sweeps for both positive and negative magnetic fields $B_z$, starting from zero to near the critical field, as shown in \figref{fig:2}(d). The curves for both signs of $B_z$ overlap to a very high degree, indicating that the polarity of the field does not influence the SDE response. The extracted values of $I_{c+}$ and $\left| I_{c-} \right|$ decrease with increasing field, as one would expect since $B_z$ weakens superconductivity. The SDE is also clearly visible in the differential resistance (dV/dI) mapping as a function of bias current and magnetic field in \figref{fig:2}(e), which again shows a clear asymmetry in positive and negative bias currents. 
Figure \ref{fig:2}(f) presents the variation of the diode efficiency as a function of the out-of-plane magnetic field $B_z$ for temperatures ranging from 2.5 K to 6.5 K. As the field increases in either direction, $\eta$ increases sharply, signifying a pronounced enhancement of the SDE under the magnetic field. The curves display a symmetric V-shaped dependence at lower temperatures. This temperature and field-dependent behavior underscores the complex interplay between superconductivity, magnetism, vortex dynamics, and nonreciprocal transport mechanisms in the system.

To compare with the SDE stabilized by the $T_2$ phase, we also measured the I-V characteristics of the $T_2$-Mn$_3$Pt/Nb heterostructure, in the absence and presence of the magnetic field. In many regards, the behavior (shown in \figref{fig:3}) is qualitatively similar to $T_1$-Mn$_3$Pt/Nb, including a non-vanishing zero-field SDE that can be enhanced upon applying $B_z\neq 0$. As before, the zero-field SDE is  expected by symmetry \cite{banerjee2024altermagnetic,PartnerTheoryPaper} and provides direct evidence of the time-reversal symmetry breaking impact of the non-collinear altermagnetic order on the superconductor. However, the efficiency, see \figref{fig:3}(b,f), of the SDE is nearly one order of magnitude smaller compared to that observed in $T_1$-Mn$_3$Pt/Nb.

\section{Discussion}
One relevant question is the reason for the striking difference we find in the diode efficiency for the $T_1$ and $T_2$ phases. As already discussed above, the key symmetry difference is that only the $T_2$ phase preserves the $\sigma_v^s$ symmetry, while it is broken in the $T_1$ state. Consequently, only the latter is chiral and exhibits a finite Berry curvature, which can explain the difference in the Hall response in \figref{fig:1}(d,e). However, in both cases, the symmetry requirements for the zero-field SDE are fulfilled and also the finite canting in the $T_1$ state is not required. One possible explanation for the difference could simply be the fact that the efficiency $\eta$ crucially depends on the direction of the current flow (with zeros pinned along different high-symmetry directions for the two phases \cite{PartnerTheoryPaper}). With transport only measured along one fixed direction, this could lead to significant differences in the observed $\eta$. Naturally, sample-to-sample variations or different coupling matrix elements across the interface could be additional sources affecting the efficiency. In this context, future angle-resolved transport measurements \cite{SunbeamSetup} would provide crucial insights.

Another noteworthy observation is that, for both magnetic orders, $\eta$ can be further enhanced by application of a magnetic field $B_z$ and is an approximately even function of $B_z$. In particular, as opposed to other ferromagnetic zero-field SDE experiments \cite{Lin2022,FerromagneticProximity}, we do not observe a switching of the sign of $\eta$ (or hysteresis), at least within the field ranges in \figref{fig:2}(f) and \figref{fig:3}(f). This is most likely related to the altermagnetic nature of the magnetic order parameter, which does not couple linearly to the applied field (not at all for $T_2$, due to $\sigma_v^s$, and the coupling requires spin-orbit coupling for $T_1$ and is thus likely small, just like the canting). One natural explanation for the increase of $\eta$ with $B_z$ could thus be its detrimental impact on superconductivity, which becomes, in turn, more sensitive to the altermagnetic order parameter (see also \cite{PartnerTheoryPaper}). Of course, the magnetic field also likely increases the relevance of vortices and their dynamics, which also crucially affects the SDE, for instance as a result of asymmetric vortex pinning see, e.g., \cite{Bulaevskii2011, Gutfreund2023,PhysRevB.49.9244}. In fact, the relevance of vortices to the SDE is also indicated by the increase of $\eta$ with temperature since vortex dynamics might become a more prominent contribution to the transport properties upon approaching the critical temperature. 

In summary, our observation of a SDE in $T_1$-Mn$_3$Pt/Nb and $T_2$-Mn$_3$Pt/Nb heterostructures provides clear evidence of the impact of the magnetic order in Mn$_3$Pt on the superconducting state of Nb, despite the (nearly) compensated nature of the $T_2$ ($T_1$) textures. 
This 
highlights the huge potential of heterostructures consisting of complex magnets and superconductors to stabilize and study altermagnetic superconductivity. 
Our findings further shed light on the possible microscopic origins of superconducting nonreciprocity and underscore the potential of non-collinear altermagnets as a platform for tunable superconducting diodes for energy-efficient superconducting electronics without large net magnetizations.

\begin{acknowledgments}
This work is supported by ANRF (formerly SERB) Core Research Grant (CRG/2023/008193), Government of India. S.S. and S.M. acknowledge Sadat Riyaz and Rahul Mishra (CARE, IIT Delhi) for helping in sample fabrication. 
M.S.S., C.S., and S.M. acknowledge important discussions with J.~Shabani.
S.M. acknowledges Subhro Bhattacharjee (ICTS) for fruitful discussions. M.S.S.~acknowledges discussions with R.~Fernandes. S.S. and S.M. acknowledge Indraneel Sinha for initial standardization of sample growth. S.S. and S.M. acknowledge the Department of Physics and the Central Research Facility at IIT Delhi for various sample characterization facility. C.S. acknowledges support from the Louisiana Board of Regents.
\end{acknowledgments}

\bibliography{references}

\end{document}